\documentstyle[epsfig]{article}
\input amssym.def
\input amssym.tex
\newcommand{\psbild}[5]
{\par
 \begin{figure}[#1]
 \begin{center}
 \begin{minipage}{0cm} \end{minipage}
 \begin{minipage}{#3}
 \refstepcounter{figure}\label{#2}
 \epsfxsize=#3
 \epsffile{#4}
 \end{minipage}
 \end{center}
 \hfill
 \begin{minipage}{0cm} \end{minipage}
 \begin{center}
 \parbox{15cm}{\baselineskip11pt{\rm Figure {#2}}: {#5}}
 \end{center}
 \end{figure}}
\newcommand{\resection}[1]{\setcounter{equation}{0}\section{#1}}

\newcommand{\beq}{\begin{equation}}
\newcommand{\eeq}{\end{equation}}
\newcommand{\bdm}{\begin{displaymath}}
\newcommand{\edm}{\end{displaymath}}
\newcommand{\bea}{\begin{eqnarray}}
\newcommand{\eea}{\end{eqnarray}}

\begin{document}
\setcounter{page}{0}
\topmargin 0pt
\newpage
\setcounter{page}{0}
\begin{titlepage}
\vspace{0.5cm}
\begin{flushright}{\bf }
\end{flushright}
\vspace{2.0cm}
\begin{center}
{\Large {Inverse scattering and solitons in $A_{n-1}$ affine Toda field 
theories II}}
\\
\vspace{1cm}
{\large Edwin J. Beggs$^*$ and Peter R. Johnson$^\dagger$}
\vspace{1.0cm}

{$*$ \em Department of Mathmatics, \\ University of Wales, Swansea, 
\\ Singleton Park, \\Swansea, \\SA2 8PP, \\UK}
\vspace{1.0cm}

{$\dagger$ \em 
Albert Einstein Institut,\\ Max-Planck-Institut f\"{u}r
Gravitationsphysik,\\ Schlaatzweg 1,\\  D-14473 Potsdam,\\ Germany }
\end{center}
\vspace{1cm}
\setcounter{footnote}{0}\baselineskip=12pt
\begin{center}{\bf Abstract} \\ \end{center}

New single soliton solutions to the affine Toda field theories are constructed,
exhibiting previously unobserved topological charges. This goes some
 of the way in filling the weights of the fundamental
representations, but nevertheless holes in the 
representations remain. 
We use the group doublecross product form of the
 inverse scattering method, and restrict ourselves to the rank one solutions.
\end{titlepage}\baselineskip=12pt
\resection{Introduction}
This paper should be considered as the sequel to our earlier paper \cite{BJ},
where some single soliton solutions to the $A_{n-1}$ affine Toda field theories
were computed by studying the space-time evolution of the residues of 
simple poles in the underlying loop group, which is a variation of the
inverse scattering method.  In the original paper
this was carried out with the aim of finding new single soliton 
solutions with new
topological charges which would fill some of the missing weights in the
fundamental representations, and thereby go some way towards confirming 
the affine quantum group symmetry of the model. Filling all the weights
is needed to justify the S-matrices calculated in \cite{J}.
The reader can refer to McGhee \cite{McGhee} for the calculation of the 
previously known topological charges. 
The attempt in the original paper failed. 
Although extra solutions were found, 
it was discovered that they simply represented modes of oscillation 
around the standard solutions, and were not fundamentally new, and certainly
did not yield any new topological charge sectors. It was
also seen how these solutions could be obtained by restricting multi-soliton
solutions. In this paper, we shall exhibit some new solutions which
lead to new topological charges, and which have the correct weights 
which lie in the appropriate fundamental representations.  
It will also be easy enough to see that the new solitons have the correct 
masses, corresponding to the multiplet which they belong to, and that they
scatter classically in the correct way, i.e. they have the same
time delays \cite{FJKO} when they scatter with other solitons.
However not all the missing charges are found by this construction,
in particular no new charges are found for $A_3$.

We shall also show that
the new solutions can be obtained by restricting more 
complicated previously known multi-soliton solutions, so that the new 
solutions could have been found without using inverse scattering.
However, it has become clear that inverse scattering has the advantage of
giving a unified approach to the possible `variations on the theme of
a single soliton'. While these solutions can be derived from multi-solitons,
it is not at all obvious to see how to restrict the momenta of the 
multi-solitons to effectively give a single soliton.

We use a loop group, defined in \cite{BJ}, which consists of matrix valued 
meromorphic functions on $\Bbb C\setminus\{0\}$ satisfying a certain
symmetry condition. The affine Toda equation is solved by reversing the
order of factorisation between an analytic element and a meromorphic
element regular at $0$ and $\infty$. However in the earlier paper
we used a formulation of the meromorphic function which was not the most 
general.

In this paper, we modify the procedure used in \cite{BJ}, 
and for the rank one case manage to 
determine the space-time evolution of the kernel of the residue
of the meromorphic function independently
to that of the image, where previously a special projection for the
residue was chosen.  The projection
was originally chosen so that the inverse loops had a particularly
attractive form, which was necessary because we could not determine
the space-time evolution of the kernel of the residue, only the image. 
In fact with this projection, the kernel was fixed completely by the image,
and this allowed us to side-step the issue about the kernel, without meeting
the problem head on.

\resection{Preliminaries}
We recall the setup and notation used in \cite{BJ}.
The affine Toda field equations for the affine algebra $\hat{g}$
follow from the zero-curvature condition
\beq
[\partial_++A_+,\partial_-+A_-]=0\label{eq: zcc},\eeq
where $A_{\pm}$ are given, in terms of the spectral parameter $\lambda$, by
\beq
A_{\pm}\ =\ \pm\frac12\beta\partial_\pm(u.H)\ \pm\ \lambda^{\pm 1}\mu
e^{\pm\frac12\beta u.H}E_{\pm 1}e^{\mp\frac12\beta u.H},
\label{eq: As} \eeq
where
$x_{\pm}=t\pm x$, and $H$ is the Cartan-subalgebra of $g$. 
In a basis where $H$ is 
diagonal  and for $g=A_{n-1}$, in the vector representation
the $n\times n$ matrices $E_{\pm 1}$ are defined to be 
$$E_{+1}={E_{-1}}^\dagger=\pmatrix{0 & 1 & 0 & \cdots & & 0 \cr & 0 & 1 & 
0 & \cdots & \cr
& & 0 & 1 & 0 & \cdots \cr \vdots & & &\ddots & \ddots & \cr
0 & 0 & \cdots & &0& 1 \cr
1 & 0 & \cdots & & & 0 }.$$

We are interested in solutions $\Phi(\lambda)$ to the linear system: 
\beq \partial_\pm\Phi=\Phi A_\pm. \label{eq: lin_sys}\eeq
The simple vacuum solutions are given by $u$ constant with 
$e^{\beta u.H/2}$ commuting with $E_{+1}$ and $E_{-1}$. In this case we define
\beq A_+=J(\lambda)=\mu\lambda E_{+1},\qquad
A_-=K(\lambda)=-\mu\lambda^{-1}E_{-1}\label{eq: JK_def}.\eeq
The equations 
$$\partial_+\Phi_0=\Phi_0J\quad {\rm and}\quad \partial_-\Phi_0=\Phi_0K, $$
have the simple exponential solution 
$$\Phi_0=Ce^{Jx_+ + Kx_-},$$
and if we subtract off this solution from $\Phi$ by setting
$$\phi=\Phi_0^{-1}\Phi,$$
we see that $\phi$ obeys the equations
\beq \partial_+\phi=\phi A_+ - J\phi\qquad {\rm and}\quad
\partial_-\phi=\phi A_- - K\phi \label{eq: final_lin_system}. \eeq
To construct soliton solutions, we suppose that $\phi$ is a
meromorphic function of $\lambda$ which is regular at $0$ and $\infty$.
The classical `vacuum' element $a(x,t)$ is defined by
$a(x,t)=e^{-J(\lambda)x_+-K(\lambda)x_-}$, 
and is an analytic function of $\lambda\in \Bbb C\setminus\{0\}$.
In \cite{BJ}, the group factorisation result
\beq a(x,t).\phi(\lambda,0,0)=\phi(\lambda,x,t).b,\label{eq: fac} \eeq
where $b(x,t)$ is another analytic function of
$\lambda$, is established.

We also define the diagonal matrix
\beq
U=\pmatrix{1 & & & & \cr
             & \omega^{-1}& & & \cr
             & & \omega^{-2}& & \cr
             & & & \ddots & \cr
             & & & & \omega^{-(n-1)}},\eeq
where we recall that $\omega=e^{2\pi i\over n}$. With these
definitions for $U$ and $E_{\pm 1}$, it is easy to check the relations
\beq
UE_{\pm 1}U^\dagger=\omega^{\pm 1}E_{\pm 1}.\eeq
It follows that we can consider solutions $\phi(\lambda)$ that 
satisfy the condition \beq U\phi(\lambda)U^\dagger\ =\ \phi(\omega\lambda).
\label{eq: important}\eeq

\resection{The form of the meromorphic loops}
We begin by considering a meromorphic  $\phi$ which satisfies the
symmetry condition (\ref{eq: important}).
 If $\phi$ has a pole at $\alpha$, 
then by (\ref{eq: important}) it also has poles
at $\{\omega\alpha, \omega^2\alpha, \ldots, \omega^{n-1}\alpha\}$. 
The most general form
for $\phi$ satisfying (\ref{eq: important}) with a 
single set of poles is
\beq
\phi(\lambda)=\Bigl({\lambda \zeta\over\lambda-\alpha} +
{\lambda U \zeta U^\dagger\over\lambda-\omega\alpha} + \cdots +
{\lambda U^{n-1} \zeta{U^\dagger}^{n-1}\over\lambda-\omega^{n-1}\alpha}
+ k \Bigr)e^{-\beta u.H/2},\label{eq: phi1}\eeq where
\beq k=\pmatrix{k_1^n & 0 &  \cdots & 0 \cr
0 & k_2^n & \cdots & 0 \cr  
0 & 0 & \ddots & 0 \cr
 0 & 0 & \cdots & k_n^n}.\eeq
The factor of $e^{-\beta u.H/2}$ is introduced for convenience, 
as we shall see
 in a moment. By summing geometric series we write, using
$z={\lambda\over\alpha}$,
\beq {\lambda \zeta\over\lambda-\alpha} +
{\lambda U \zeta U^\dagger\over\lambda-\omega\alpha} + \cdots +
{\lambda U^{n-1} \zeta{U^\dagger}^{n-1}\over\lambda-\omega^{n-1}\alpha}
={nz\over z^n-1}
\pmatrix{z^{n-1}\zeta_{11} & \zeta_{12} & 
z \zeta_{13} & \cdots & z^{n-2}\zeta_{1n}  \cr
z^{n-2}\zeta_{21} & z^{n-1}\zeta_{22} &
 \zeta_{23} & \cdots & z^{n-3}\zeta_{2n} \cr 
\vdots & & & & \vdots\cr
z\zeta_{n-1 1} & z^2\zeta_{n-1 2} & z^3\zeta_{n-1 3} & \cdots &\zeta_{n-1 n}
 \cr
 \zeta_{n1} & z\zeta_{n2} & z^2\zeta_{n3} & \cdots & z^{n-1}\zeta_{nn} }.\eeq
Now we must discuss how to fix the diagonal matrix $k$. As in
equation (6.1) of \cite{BJ}
we impose the  condition
\beq \phi(\infty)=1.e^{-\beta u.H/2}, \label{eq: c_cond}\eeq
or equivalently \beq k_i^n=1-n a_{ii}, \qquad i=1,\ldots n.
\label{eq: c_cond2}\eeq

In this paper we will only discuss the case where the matrix residue
$\zeta$ is of rank one.
The most general rank one matrix can be written
\beq
\zeta=\pmatrix{v_1 & 0& 0& \cdots & 0\cr
 v_2 & 0& 0 &\cdots & 0\cr \vdots & 0 & 0& \cdots & 0\cr v_n & 0& \cdots & 0 & 0\cr}
\pmatrix{ w_1
 & w_2 & \cdots & w_n\cr 0 & 0 & \cdots & 0\cr 0 & 0 & \vdots & 0\cr
0 & 0 & \cdots & 0}.\eeq
Now we can calculate
\beq
{\rm det}(\phi e^{\beta u.H/2})={\lambda^n-(k_1 k_2\ldots k_n \alpha)^n\over
\lambda^n-\alpha^n} .\label{eq: dett} \eeq
Now we examine $\phi(\lambda)^{-1}$, which is required in subsequent sections.
From the determinant (\ref{eq: dett}),
$\phi(\lambda)^{-1}$ only has one set of poles at
$\lambda=\gamma,\omega\gamma,\ldots,\omega^{n-1}\gamma$, 
for $\gamma=\tau\alpha$,
where for convenience we set
 $\tau=k_1 k_2\ldots k_n$.
The condition $U\phi(\lambda)U^\dagger=\phi(\omega\lambda)$ implies
$U\phi(\lambda)^{-1}U^\dagger=\phi(\omega\lambda)^{-1}$, so
$\phi(\lambda)^{-1}$ can be written in the same form as
$\phi(\lambda)$, i.e.\ equation (\ref{eq: phi1}), but with different
matrices instead of $\zeta$ and $k$. It is possible to show that \beq
\phi(\lambda)^{-1}=e^{\beta u.H/2}\Bigl({\lambda \xi\over\lambda-\gamma} +
{\lambda U \xi U^\dagger\over\lambda-\omega\gamma} + \cdots + {\lambda
U^{n-1} \xi{U^\dagger}^{n-1}\over\lambda-\omega^{n-1}\gamma} + k^{-1}
\Bigr),\label{eq: inversion} \eeq
where $\xi$ is given explicitly by
$$\xi=\Big({\tau^n\over k_1^n}\Big)\pmatrix{v_1 & 0& 0 &\cdots & 0\cr
 v_2 (\tau^{-1}k_1^n) & 0& 0 &\cdots & 0\cr 
v_3(\tau^{-2}k_1^n k_2^n) & 0& 0 &\vdots & 0\cr
v_4(\tau^{-3}k_1^n k_2^n k_3^n)  & 0& 0 &\vdots & 0\cr
\vdots & 0 & 0 &\vdots & 0 \cr
 v_n(\tau^{-(n-1)}k_1^n k_2^n\cdots k_{n-1}^n) & 0& 0 & \cdots& 0\cr}
\qquad\qquad\qquad\qquad\qquad\qquad\qquad$$
\bea \times
\pmatrix{ w_1  & w_2(\tau k_2^{-n}) & w_3 (\tau^2 k_2^{-n} k_3^{-n}) &
w_4 (\tau^3 k_2^{-n} k_3^{-n}k_4^{-n})& \cdots & w_n (\tau^{n-1} k_2^{-n}
k_3^{-n}\cdots k_n^{-n})\cr 0 & 0 & 0 & 0 &\cdots & 0\cr \vdots & 0 & 0 & 0 &\cdots &  \vdots\cr
0 & 0 & 0 &0 &\cdots & 0}.\label{eq: bigequn}
\eea

\resection{Reconstruction of the meromorphic loop}
We shall take the dot product of vectors to be
$u.v=u_1v_1+\dots +u_nv_n$ in terms of components. This omits the more usual
complex conjugate, but we shall find this form convenient for constructing
matrices.

It is our task in this section to reconstruct the
meromorphic function $\phi$ from the data
$$V={\rm Im\ }\Bigl({\rm res\ }\phi(\lambda)\Bigl|_{\lambda=\alpha}\Bigr)
\quad{\rm and}\quad 
W={\rm Ker\ }\Bigl({\rm res\ }\phi(\lambda)^{-1}
\Bigl|_{ \lambda=\gamma}\Bigr)\ ,$$
provided the extra complex number $\tau$ is given.
The subspace $W$ is $n-1$ dimensional, and we shall describe it by
the vector $s$ which is perpendicular to $W$, i.e.
$s.m$=0 for all $m\in W$. The vector $s$ is only determined up to scalar
multiple, but we can read off the ratios of its components
from equation (\ref{eq: bigequn}) as
\bea
{w_2\over w_1} {\tau\over k_2^n}&=&{s_2\over s_1} \cr\cr
{w_3\over w_1} {\tau^2\over k_2^n k_3^n}&=&{s_3\over s_1} \cr\cr
{w_4\over w_1} {\tau^3\over k_2^n k_3^n k_4^n}&=&{s_4\over s_1} \cr\cr
\vdots &=& \vdots \cr\cr
{w_n\over w_1} {\tau^{n-1}\over k_2^n k_3^n\cdots k_n^n}&=&{s_n\over s_1}
\label{eq: biglist}
\eea
 From the  condition (\ref{eq: c_cond2})
we could write each $k_i$ in terms of $v$ and $w$. However
we have no direct knowledge of the vector $w$, we just know $v$ and $s$
up to scalar multiples. We combine (\ref{eq: c_cond2})
with (\ref{eq: biglist}), and write $p_i=n v_is_i$, to get the
following results:
\bea k_2^n&=&1-n w_2 v_2\cr\cr\cr
&=&1-{p_2\over\tau}{w_1\over s_1}k_2^n.\eea
Thus
$$k_2^n={1\over 1+ {p_2\over\tau}{w_1\over s_1}},$$
Now $$k_3^n=1-{p_3\over\tau^2}{w_1\over s_1}k_2^n k_3^n,$$
so $$k_3^n={{s_1\over w_1}+ {p_2\over\tau}\over
{s_1\over w_1} + {p_2\over\tau} + {p_3\over\tau^2}},$$
and continuing in this way, we see that
\beq 
k_i^n = {{s_1\over w_1} + {p_2\over\tau} + {p_3\over\tau^2} 
+ \cdots + {p_{i-1}\over\tau^{i-2}}\over 
{s_1\over w_1} + {p_2\over\tau} + {p_3\over\tau^2} 
+ \cdots + {p_{i-1}\over\tau^{i-2}} + 
{p_{i}\over\tau^{i-1}}},\label{eq: ks}\eeq
for $i=2,\cdots n$.
The  condition (\ref{eq: c_cond2}) also gives
\beq k_1^n=1-p_1 {w_1\over s_1}\label{eq: k1}\eeq 
If we use the fact that 
$$k_1^n k_2^n \cdots k_n^n = \tau^n,$$
then
$${({s_1\over w_1}-p_1)\over {s_1\over w_1} + {p_2\over\tau} +
{p_3\over\tau^2} + \cdots + {p_n\over\tau^{n-1}}}=\tau^n\ ,$$
resulting in
$${s_1\over w_1} = {p_1 + p_n \tau + p_{n-1}\tau^2 + p_{n-2}\tau^3 + 
\cdots + p_2\tau^{n-1}\over 1-\tau^n}.$$
Substitution back into (\ref{eq: k1}) gives
$$k_1^n={\tau(p_n + p_{n-1} \tau + \cdots + p_2\tau^{n-2} + p_1\tau^{n-1})
\over
p_1 + p_n \tau + p_{n-1} \tau^2 + \cdots + p_2\tau^{n-1}},$$
and for general $i$, substituting into (\ref{eq: ks}) and 
with the understanding of taking $j$ mod $n$ in 
$p_j$:
$$k_i^n={\tau(p_{n-1+i} + p_{n-2+i} \tau + \cdots + p_{1+i}\tau^{n-2} + 
p_i\tau^{n-1})\over
p_i + p_{i-1} \tau + p_{i-2} \tau^2 + \cdots + p_{i-n+1}\tau^{n-1}}.$$
We have thus reconstructed the $k_i$ from the vectors $v$ and $s$, together
with $\tau$.

\resection{The rank one solution} 
We begin by specifying the initial meromorphic loop $\phi_0$
with poles at $\{\alpha,\omega\alpha,\ldots,\omega^{n-1}\alpha\}$. 
The factorisation (\ref{eq: fac}) solves the space-time evolution
of $\phi$, given some initial $\phi_0$. Thus
$a.\phi_0=\phi.b$, where $a$ and $b$ are analytic on $\Bbb C\setminus\{0\}$,
and we see that the pole position $\alpha$ of $\phi$ is independent of 
space-time.
The inverse
$\phi_0^{-1}$ has poles at 
$\{\gamma,\omega\gamma,\ldots,\omega^{n-1}\gamma\}$.
From the factorisation $\phi_0^{-1}.a^{-1}=b^{-1}.\phi^{-1}$
we see that $\gamma$ for $\phi^{-1}$ also does not depend on space-time,
and from this we deduce that $\tau$ must be a constant.

As $\phi_0$ is specified, we know
$$V_0={\rm Im\ }\Bigl({\rm res\ }\phi_0(\lambda)\Bigl|_{\lambda=\alpha}\Bigr)
\quad{\rm and}\quad 
W_0={\rm Ker\ }\Bigl({\rm res\ }\phi_0(\lambda)^{-1}
\Bigl|_{ \lambda=\gamma}\Bigr)\ .$$
For the factorisation $a.\phi_0=\phi.b$, we take the residues
at $\lambda=\alpha$ of both sides of the equation. Since $b(\alpha)$ is 
invertible, by comparing images it follows that
$$
V\ =\ 
{\rm Im\ }\Bigl({\rm res\ }\phi(\lambda)\Bigl|_{\lambda=\alpha}\Bigr)
\ =\ 
a(\alpha) V_0\ .
$$
If we invert the factorisation to get 
$\phi_0^{-1}.a^{-1}=b^{-1}.\phi^{-1}$, taking the residues
of both sides at $\lambda=\gamma$, and applying a vector $m$ to both sides,
we get
$$
\Bigl({\rm res\ }\phi_0(\lambda)^{-1}
\Bigl|_{ \lambda=\gamma}\Bigr)a(\gamma)^{-1}m\ =\ 
b(\gamma)^{-1}
\Bigl({\rm res\ }\phi(\lambda)^{-1}
\Bigl|_{ \lambda=\gamma}\Bigr)m\ .
$$
From this it follows that
$$
a(\gamma)^{-1}m\in 
{\rm Ker\ }\Bigl({\rm res\ }\phi_0(\lambda)^{-1}
\Bigl|_{ \lambda=\gamma}\Bigr)
\quad\mbox{\rm if and only if}\quad
m\in 
{\rm Ker\ }\Bigl({\rm res\ }\phi(\lambda)^{-1}
\Bigl|_{ \lambda=\gamma}\Bigr)\ ,
$$
thus
$$
W\ =\ {\rm Ker\ }\Bigl({\rm res\ }\phi(\lambda)^{-1}
\Bigl|_{ \lambda=\gamma}\Bigr)\ =\ a(\gamma)W_0\ .
$$

We will take the one dimensional subspace $V_0$ to 
be spanned by the vector $v_0$,
and then $V$ will be spanned by $v=a(\alpha)v_0$. The $n-1$ dimensional
subspace $W_0$ will be specified by its perpendicular vector $s_0$. Then
the perpendicular vector $s$ describing $W$ is defined by
$s=(a(\gamma)^T)^{-1}s_0$.

As $\lambda=0$ is a regular point of $\phi(\lambda)$, from the
 linear system (\ref{eq: final_lin_system}) we have
\beq
\phi(0)^{-1}E_{-1}\phi(0)=e^{-\beta u.H/2}E_{-1}e^{\beta u.H/2}\ .
\label{eq: regul0}\eeq
Since $\phi(0)$ is diagonal  we must have 
$${1\over\tau}\pmatrix{k_1^n & 0 &  \cdots & 0 \cr
0 & k_2^n & \cdots & 0 \cr  
0 & 0 & \ddots & 0 \cr
 0 & 0 & \cdots & k_n^n}=e^{\beta u.H}.$$
Thus to determine $e^{\beta u.H}$ we have to find the $k_i$ from our known
quantities $v$, $s$ and $\tau$, which is what we did in the last section.
Writing the fundamental weights $\{\lambda_i: i=1,\ldots,n-1\}$
in terms of the weights of the basic representation (the eigenvalues of $H$),
we get the solution:
\bea
e^{-\beta\lambda_1.u}&=&{p_1 + p_n \tau + p_{n-1} \tau^2 + \cdots +
 p_2\tau^{n-1}\over
p_n + p_{n-1} \tau + \cdots + p_2\tau^{n-2} + p_1\tau^{n-1}}\cr\cr
e^{-\beta\lambda_2.u}&=&{p_2 + p_1 \tau + p_{n} \tau^2 + \cdots +
 p_3\tau^{n-1}\over
p_n + p_{n-1} \tau + \cdots + p_2\tau^{n-2} + p_1\tau^{n-1}}\cr\cr
e^{-\beta\lambda_3.u}&=&{p_3 + p_2 \tau + p_{1} \tau^2 + \cdots +
 p_4\tau^{n-1}\over
p_n + p_{n-1} \tau + \cdots + p_2\tau^{n-2} + p_1\tau^{n-1}}\cr\cr
&\vdots&\cr\cr
e^{-\beta\lambda_{n-1}.u}&=&{p_{n-1} + p_{n-2} \tau + p_{n-3} \tau^2 +
 \cdots +
 p_n\tau^{n-1}\over
p_n + p_{n-1} \tau + \cdots + p_2\tau^{n-2} + p_1\tau^{n-1}}
\label{eq: new_solution}\eea

\resection{Discussion of the rank one result} 
We recall first the standard soliton solutions to these theories and
the rules for composing these solutions together \cite{H1}.
For \beq e^{-\beta\lambda_i.u}={\tau_i\over\tau_0},\qquad\qquad 
i=1,\ldots n-1,
\label{eq: recall} \eeq
recall that the standard
one-soliton solution of species $j$ is 
\beq 
\tau_i=1+Q\omega^{ij}W_j(\theta),\qquad\qquad i=0,\ldots n-1,
\label{eq: standard_one} \eeq
where $W_j(\theta)=e^{m_j(e^{-\theta}x_+-e^\theta x_-)}$. The constant
$m_j$ is proportional to the mass of the soliton of species $j$, which is
$m_j=2\sin{\pi j\over n}$. 
This solution can be added to some solution $\tau_i^{\rm old}$, 
already in the soliton sector of the theory, by the rule
$$\tau_i^{\rm new}=\tau_i^{\rm old} + Q\omega^{ij}W_j(\theta) + 
\hbox{\rm cross-terms},$$ 
where the cross-terms are formed by the rule:

\noindent
For any term $AW_{j_1}(\theta_1)W_{j_2}(\theta_2)\cdots W_{j_r}(\theta_r)$ in
$\tau_i^{\rm old}$, we pick up a cross-term 
$$QA\omega^{ij}X^{jj_1}(\theta-\theta_1)
X^{jj_2}(\theta-\theta_2)\cdots X^{jj_r}(\theta-\theta_r)
W_j(\theta)W_{j_1}(\theta_1)W_{j_2}(\theta_2)\cdots W_{j_r}(\theta_r).$$
The interaction coefficient $X^{jr}(\theta)$ has the definition
\beq 
X^{jr}(\theta)={(e^\theta-e^{{\pi i(j-r)\over n}})(e^\theta-e^{-{\pi i(j-r)\over n}})\over  (e^\theta-e^{{\pi i(j+r)\over n}})(e^\theta-e^{-{\pi i(j+r)\over n}})}.\label{eq: Xjr}\eeq

Thus with this rule, we can immediately write down the standard 
two-soliton solution, starting from the standard one-soliton solution above:
\beq\tau_i=1+Q_1\omega^{ij}W_j(\theta_1)+Q_2\omega^{ir}W_r(\theta_2) 
+Q_1Q_2\omega^{i(j+r)}X^{jr}(\theta_1-\theta_2)W_j(\theta_1)W_r(\theta_2).
\label{eq: two_sol}\eeq

We now consider the solution derived by the rank one inverse scattering 
method, 
equation (\ref{eq: new_solution}), and discuss how it is composed of $W$'s,
and how it can be recovered by considering multi-soliton solutions. 

Let $e_i\,:i=0,\ldots,n-1$ be the simultaneous eigenvectors of $E_{+1}$
and $E_{-1}$ (recall that \\$[E_{+1},E_{-1}]=0$)
$$e_i=\pmatrix{ 1 \cr \omega^i \cr \omega^{2i}\cr \vdots \cr
\omega^{(n-1)i}},\qquad {\rm where}\qquad E_{\pm 1}e_i=\omega^{\pm
i}e_i\,:\qquad i=0,\ldots,n-1.$$
Then 
$$a(\alpha)e_j=e^{-\alpha\mu E_{+1}x_++\alpha^{-1}\mu E_{-1}x_-}e_j=
e^{-\mu\alpha\omega^jx_++\mu\alpha^{-1}\omega^{-j}x_-}e_j.$$
We set
$$W_j=e^{-\mu\alpha(\omega^j-1)x_++\mu\alpha^{-1}(\omega^{-j}-1)x_-},$$
fixing the phase of $\alpha$ by setting $\alpha=i\omega^{-j/2}e^{-\theta}$,
 then
$$W_j(\theta)=e^{m_j(e^{-\theta}x_+-e^{\theta}x_-)},$$ 
as before.
We also introduce the set $B_j$, as in \cite{BJ}, defined as the set of all 
integers $k$ such that 
$$m_j>m_k\cos\Bigl({(j-k)\pi\over n}\Bigr).$$
Then we can define, for $k\in B_j$,
$$W_k\bigl(\theta+{(j-k)\pi i\over n}\bigr)
=e^{m_k(e^{-(\theta+{(j-k)\pi i\over n})}x_+-
e^{(\theta+{(j-k)\pi i\over n})}x_-)}\ .$$
Given the initial image vector $v_0= e_0 + \sum_{k\in B_j}Q_k e_k + 
Q_j e_j$, the image $v$ evolves as 
\beq v=e_0 + \sum_{k\in B_j}Q_k W_k\bigl(\theta+{(j-k)\pi i\over n}\bigr) e_k +
Q_j W_j(\theta)e_j.\label{eq: v_i} \eeq 
This defines so-called `right moving' modes for the standard soliton of 
species $j$. As shown in \cite{BJ}, these can be formed by composing solitons
together, and repeatedly setting $X^{kj}(\theta_k-\theta_j)=0$ for 
$k\in B_j$, by choosing values of $\theta_k$ which lie at a zero, i.e.\ 
the second zero in (\ref{eq: Xjr}).
  
We can also define `left movers', by considering the evolution of the vector
$s$, orthogonal to the kernel. Pick the initial $s_0$, as
$s_0=e_0+ \sum_{k\in B_r}\tilde{Q}_k e_k + \tilde{Q}_r e_r$, 
then recalling that
$s=(a(\tau\alpha)^{T})^{-1}s_0$,
\beq 
s=e_0+ \sum_{k\in B_r}\tilde{Q}_k W_k\bigl(\tilde{\theta}-{(r-k)\pi 
i\over n}\bigr) e_k + \tilde{Q}_r W_r(\tilde{\theta})e_r,\label{eq: s_i}\eeq
where 
\beq e^{\tilde{\theta}}=\omega^{{j+r\over 2}} \tau^{-1}e^\theta.
\label{eq: th}\eeq
These left movers were not considered in \cite{BJ}, but arise because
there are always two zeros to the interaction coefficient $X^{kr}(\theta)$,
at $\theta=\nu$ and at $\theta=-\nu$, and we take the alternative case
to that considered for the image. 

By expanding out the new solution (\ref{eq: new_solution}), it is possible
to see that it is the result of combining
a standard soliton of species $j$ with right moving modes of oscillation 
with a 
standard soliton of species $r$ with left-moving modes. 
This expansion will be done below. It is only necessary to pick
a value of $\tau$ or $\tilde{\theta}$ to ensure that the mass of the
resultant solution is what we are looking for. This is done by
again picking a zero of the interaction coefficient 
$X^{jr}(\theta-\tilde{\theta})$. Important cross terms arise
when left-movers interact with right-movers, and it is these which lead to 
the new topological charge sectors. 

The new solutions (\ref{eq: new_solution}) are given by 
$$\tau_i=p_i + p_{i-1}\tau + p_{i-2}\tau^2 + \cdots + 
p_{i+1}\tau^{n-1},\qquad i=0,\ldots n-1,$$
with the understanding that $p_0=p_n$, and the index $j$ of $p_j$ is taken
modulo $n$. We first take the simplest situation with no extra modes,
we write $W_j$ for $W_j(\theta)$, and $\tilde{W}_r$ for $W_r(\tilde{\theta})$:
$$v_i=1+Q_j\omega^{j(i-1)}W_j,$$
$$s_i=1+\tilde{Q}_r\omega^{r(i-1)}\tilde{W}_r.$$
Now $p_i=n v_i s_i$, so
\bea
\tau_i&=&(1+Q_j\omega^{j(i-1)}W_j)(1+\tilde{Q}_r\omega^{r(i-1)}\tilde{W}_r)
+\tau (1+Q_j\omega^{j(i-2)}W_j)(1+\tilde{Q}_r\omega^{r(i-2)}\tilde{W}_r)
+\cdots \cr\cr
&&+\ 
\tau^{n-1}(1+Q_j\omega^{ji}W_j)(1+\tilde{Q}_r\omega^{ri}\tilde{W}_r) \cr\cr
&=&(1+\tau+\tau^2+ \cdots +\tau^{n-1}) + Q_j\omega^{ji}W_j(\omega^{-j}
+\omega^{-2j}\tau + \cdots +\tau^{n-1})\cr\cr
&& +\ \tilde{Q}_r\omega^{ri}\tilde{W}_r(\omega^{-r}+
\omega^{-2r}\tau + \cdots +\tau^{n-1}) \cr\cr
&&+\ Q_j\tilde{Q}_r\omega^{(j+r)i}W_j\tilde{W}_r
(\omega^{-(r+j)}+\omega^{-2(r+j)}\tau + \cdots +\tau^{n-1}).\eea
We divide $\tau_i$ by $(1+\tau+\tau^2+ \cdots +\tau^{n-1})$, and 
let
 $$\tilde{Q}'_r=\tilde{Q}_r
{(\omega^{-r}+\omega^{-2r}\tau + \cdots +\tau^{n-1})\over
(1+\tau+ \cdots +\tau^{n-1})},$$
 $$Q'_j=Q_j{(\omega^{-j}+\omega^{-2j}\tau + \cdots +\tau^{n-1})\over
(1+\tau+ \cdots +\tau^{n-1})}.$$
Then $$\tau_i=1+Q'_j\omega^{ji}W_j + \tilde{Q}'_r\omega^{ri}\tilde{W}_r
+ X^{jr}(\theta-\tilde{\theta})Q'_j\tilde{Q}'_r\omega^{i(j+r)}
W_j\tilde{W}_r,$$
where we have made the identification
$$X^{jr}(\theta-\tilde{\theta})={(\omega^{-(r+j)}+\omega^{-2(r+j)}\tau +
 \cdots +\tau^{n-1})(1+\tau+\cdots + \tau^{n-1})\over
(\omega^{-j}+\omega^{-2j}\tau + \cdots +\tau^{n-1})
(\omega^{-r}+\omega^{-2r}\tau + \cdots +\tau^{n-1})}.$$
Using the formula for the sum of a geometric progression
four times on this expression shows that it is equal to the
formula (\ref{eq: Xjr}), after using (\ref{eq: th}).
Thus we have derived the two-soliton solution
(\ref{eq: two_sol}).

This shows how the interaction coefficient $X^{jr}(\theta)$ appears from 
a $W_j$ term in $v$  combining with the $W_r$ term in $s$. From this it is
easy to see how the correct interaction coefficients are picked up when other
modes in $v_i$ (\ref{eq: v_i}) (or $s_i$) combine with modes from $s_i$ 
(\ref{eq: s_i}) (or $v_i$) respectively.

It is also amusing to note that the standard breather solutions, 
for example mentioned  in \cite{OTUb}, considered as
a bound state of a soliton and anti-soliton, (and an analytic continuation
in the relative rapidity to give a total real energy and momentum),
are contained as a simple case of the solutions (\ref{eq: new_solution}).
These are recovered if no extra modes are taken, $j=\bar{r}$, 
and $\tau$ has some special values. The breathers are of course single
extended objects, so we would hope to find them in this way.

\resection{Some explicit new single solitons}
We have to be careful to choose modes so that any resultant exponential 
behaviour in $x$ does not exceed the behaviour of the final term in the 
tau function, which is  the term we have selected to provide the mass
of the soliton we are interested in, see \cite{BJ} for a discussion of this.
To illustrate how we obtain new topological charges we shall focus
on the largest fundamental representation (the middle node of the
Dynkin diagram) of the $A_n$, $n$ odd, theories.
This representation always has two previously known topological charges.
For the combined 
modes not to exceed the mass term, we must take $n\ge 5$, so the new rank
one solutions fail to solve the $A_3$ theories where the middle representation
is of dimension $6$, so that there are four missing charges.

So we consider $A_5$ in detail, this being the next simplest case, the
first case where we get something new. The middle fundamental
representation $V_3$, corresponding to the soliton of species $3$,
has dimension $20$. There are therefore $18$ missing charges.

Consider the `right movers' around a standard soliton of species $3$,
$$\tau_j^{(1)}=1+\omega^j Q_1 W_1(\theta-{2\pi i\over 6}) + \omega^{5j}Q_5
W_5(\theta+{2\pi i\over 6}) + \omega^{3j}Q_3W_3(\theta),$$
and the left-movers around a standard soliton of species $3$,
$$\tau_j^{(2)}=1+\omega^j \tilde{Q}_1 W_1(\theta+{2\pi i\over 6}) + 
\omega^{5j}\tilde{Q}_5
W_5(\theta-{2\pi i\over 6}) + \omega^{3j}\tilde{Q}_3W_3(\theta).$$
We have seen that the rank one solutions (\ref{eq: new_solution}) are
given by combining these solitons with left-moving and right-moving modes 
together.
We have set the relative rapidity of these two solitons to zero so 
that the resultant mass is the same as a species $3$ soliton, the final
term in each tau function remains, and there are no higher terms in $x$.
Thus the combined soliton, which we interpret as a single soliton, is
\bea \tau_j&=&1+ (\omega^j Q_1 + \omega^{5j}\tilde{Q}_5)W_1
(\theta-{2\pi i\over 6}) + (\omega^{5j} Q_5 + \omega^{j}\tilde{Q}_1)W_1
(\theta+{2\pi i\over 6}) \cr\cr
&&+\ X^{11}(-{4\pi i\over 6})\omega^{2j}Q_1\tilde{Q}_1W_1(\theta-
{2\pi i\over 6})
W_1(\theta+{2\pi i\over 6}) \cr\cr
&&+\  X^{55}({4\pi i\over 6})\omega^{4j}Q_5
\tilde{Q}_5W_5(\theta+{2\pi i\over 6})W_5(\theta-{2\pi i\over 6}) \cr\cr
&&+\  X^{51}(0)Q_5
\tilde{Q}_1W_5(\theta+{2\pi i\over 6})W_1(\theta+{2\pi i\over 6}) \cr\cr
&&+\  X^{15}(0)Q_1
\tilde{Q}_5W_1(\theta-{2\pi i\over 6})W_5(\theta-{2\pi i\over 6}) + 
\omega^{3j}(Q_3+\tilde{Q}_3)W_3(\theta)\ . \label{eq: explicit}
\eea
We know that
\bea X^{11}(-{4\pi i\over 6})=X^{55}({4\pi i\over 6})&=&{3\over 2}
\quad{\rm and}\quad
X^{15}(0)=X^{51}(0)={3\over 4}\ .\eea
From the formula $m_j=\sin({\pi j\over n})$, we calculate
$$W_1(\theta)=W_5(\theta)=e^{{1\over 2}(e^{-\theta}x_+-e^\theta x_-)}\quad
{\rm and}\quad
W_3(\theta)=e^{(e^{-\theta}x_+-e^\theta x_-)}\ ,$$
so $W_3(0)=e^{2x}$.
Also
\bea 
W_1(-{2\pi i\over 6})&=&
e^{{1\over 2}(e^{2\pi i\over 6}x_+-e^{-{2\pi i\over 6}}
x_-)}=
e^{{1\over 2}x+ i{\sqrt{3}\over 2}t}, \eea
and
$$W_1({2\pi i\over 6})=e^{{1\over 2}x- i{\sqrt{3}\over 2}t}\ .$$
For $\theta=0$, when we are working in the rest-frame of the soliton, 
equation (\ref{eq: explicit}) becomes
\bea\tau_j&=&1 +  (\omega^j Q_1 + \omega^{5j}\tilde{Q}_5)
e^{{1\over 2}x+ i{\sqrt{3}\over 2}t}
+ (\omega^{5j} Q_5 + \omega^{j}\tilde{Q}_1)
e^{{1\over 2}x- i{\sqrt{3}\over 2}t}
\cr\cr
&+&{3\over 2}\Bigl\{\omega^{2j}Q_1\tilde{Q}_1 + \omega^{4j}Q_5\tilde{Q}_5
+{1\over 2}\Bigl( Q_1\tilde{Q}_5 e^{i\sqrt{3}t} + 
Q_5\tilde{Q}_1 e^{-i\sqrt{3}t}\Bigr)\Bigr\}e^x + 
\omega^{3j}(Q'_3)e^{2x}.\label{eq: explicit2}
\eea 
Notice that as promised the intermediate term in curly brackets has
an exponential behaviour $e^x$ which does not beat the last term 
$e^{2x}$ at large $x$, thus this solution indeed has a mass equal to that
of the solitons of species $3$. If the construction were repeated
for the middle soliton of the $A_3$ theories, then this intermediate
term would have an exponential behaviour exactly equal to the final 
term. Given the time dependence in the intermediate term, this is not allowed.

\noindent {\bf Structure of the singularities}

We graphically examine the behaviour of  $\tau_j$, $j=0,\ldots,5$, as 
we vary the complex parameters $Q_1,\tilde{Q}_1, Q_5, \tilde{Q}_5, Q'_3$.
We seek values where the solutions $\tau_j$ are free from zeros, so that
the field $u$ (recall equation (\ref{eq: recall}))
is free from singularities, for all $x\in{\Bbb R}$
and $t\in{\Bbb R}$.

Consider the equation $\tau_j=0$. To simplify this equation
we partially split the $x$ and $t$ dependence by rearranging
it as
$$e^{-{x\over 2}} 
+{3\over 2}\Bigl\{\omega^{2j}Q_1\tilde{Q}_1 + \omega^{4j}Q_5\tilde{Q}_5
+{1\over 2}\Bigl( Q_1\tilde{Q}_5 e^{i\sqrt{3}t} + Q_5\tilde{Q}_1
 e^{-i\sqrt{3}t}\Bigr)\Bigr\}e^{{x\over 2}} + 
\omega^{3j}(Q'_3)e^{{3x\over 2}}=$$
\beq\qquad\qquad\qquad\qquad-(\omega^j Q_1 + \omega^{5j}\tilde{Q}_5)
e^{i{\sqrt{3}\over 2}t}
-(\omega^{5j} Q_5 + \omega^{j}\tilde{Q}_1)e^{-i{\sqrt{3}\over 2}t}.
\label{eq: left_right}\eeq
We plot the left and right hand sides of this equation,
for all $x\in{\Bbb R}$, and for all $t\in{\Bbb R}$.
 The right-hand
side will always be an ellipse, in fact we will mostly consider the
degenerate case when this ellipse is a line or a point.  The
$t$ dependence of the left hand-side complicates matters, and we will have
to pick certain values of $t$, i.e `extremal' values, and plot curves
for all $x\in{\Bbb R}$, these curves  come from and leave infinity.
From the graph of the left-hand side it will be a simple matter to read off 
the winding numbers around the origin of the tau functions and thus 
compute the topological charges. We also remark that if we set
$Q_1,\tilde{Q}_1, Q_5, \tilde{Q}_5$ to zero, we recover the known
solution, equation (\ref{eq: standard_one}), for a soliton of species $3$.
We aim to tune these parameters to enter a new topological charge
sector, so we must increase $Q_1,\tilde{Q}_1, Q_5, \tilde{Q}_5$ from zero 
through a singular region into  a region free from singularities. The
winding number will then  automatically be different from that previously 
known.

It makes
sense to do this tuning in a systematic manner, so we pick a particular
tau function which we want to tune into a new charge sector, $\tau_0$
say. We have to tune the curve given by the left-hand side through
and past the ellipse centred at the origin, given by the right-hand side.
This could be difficult to do since the size of the perturbation 
given by the term in curly brackets  depends on the size of the ellipse:
the ellipse can increase in size as we increase the perturbation.
Therefore we choose values so that the ellipse for this tau function collapses
to a point at the origin, i.e $Q_1=-\tilde{Q}_5$, $Q_5= - \tilde{Q}_1$,
thus making it much easier to tune the curve through the ellipse.
We then do further tuning, within this parameter space,
to make sure that the other tau functions,
$\tau_j: j=1,\ldots 5$, are free from zeroes. 
We will also expect that
the other tau functions will not be tuned into new topological charge
sectors, but this has to be seen explicitly in the graphs.
We can then repeat the whole construction by focussing in turn on 
$\tau_1$, $\tau_2$, etc, instead of $\tau_0$.

It turns out that the tuning required is rather delicate, so we shall
present the result as a {\it fait accompli}, exhibiting precise
values of  $Q_1,\tilde{Q}_1, Q_5, \tilde{Q}_5$, and $Q_3'$ which yield
a new sector. We will not attempt to prove statements about
completeness, that is whether all possible charge sectors have been found.

These precise values are 
(noting that $Q_1=-\tilde{Q}_5$, $Q_5= - \tilde{Q}_1$)
 $$Q_1=\tilde{Q}_1=\frac32 e^{-i\pi{41\over 84}}\quad{\rm and}\quad
Q'_3=e^{{i\pi\over 6}}.$$

We discuss $\tau_0$, which has been selected to give a new
topological charge, first. We check that $\tau_0$ is free from zeroes. 
For $j=0$ equation (\ref{eq: left_right}) becomes
$$
e^{-x/2}\Big(
1+ \frac 32 e^x\big(2-\cos(\sqrt 3 t)
\big)Q_1^2+Q_3'e^{2x}\Big)\ =\ 0\ ,
$$
and if we substitute $p=2-\cos(\sqrt 3 t)$ we get
\beq
e^{-x/2}\Big(
1\ +\ \Big(\frac32\Big)^3 pe^{-41 i\pi/42}e^x\ +\ e^{i\pi/6}e^{2x}
\Big)\ =\ 0\ .
\label{eq: subpppp}
\eeq
The imaginary part of 
(\ref{eq: subpppp})
 is
$$
e^{-x/2}\Big(-\Big(\frac32\Big)^3 p\sin\Big(\frac{41\pi}{42}\Big)e^x\ +\ 
\frac12 e^{2x}\Big)\ =\ 0\ ,
$$
which can be solved for $e^x$ to give
$$
e^x\ =\ 2p\Big(\frac32\Big)^3\sin\Big(\frac{41\pi}{42}\Big)\ .
$$
The real part of  (\ref{eq: subpppp}) is
$$
e^{-x/2}\Big(1\ +\ p\Big(\frac32\Big)^3\cos\Big(\frac{41\pi}{42}\Big)e^x\ +\ 
\frac{\sqrt 3}2 e^{2x}\Big),$$
and substituting the previous expression for $e^x$ in this gives
$$
e^{-x/2}\Big(
1\ +\ p^2\Big(\frac32\Big)^6\Big( 2\cos\Big(\frac{41\pi}{42}\Big)
\sin\Big(\frac{41\pi}{42}\Big)+2\sqrt 3 \sin^2\Big(\frac{41\pi}{42}\Big)
\Big)\Big)\ ,$$
which is the co-ordinate on the real axis where the curve passes through that 
axis. 
In the permitted region for $p$, 
$1\le p\le 3$, this value is always strictly negative and  we conclude 
that the curve $\tau_0$, parametrised by $x$,  never passes through the
origin at any time $t$.

The following two graphs ($1a$) and ($1b$) show the left-hand side of
equation (\ref{eq: left_right}) with $j=0$
at the two possible extremal values in time $t$. As discussed beforehand,
the right-hand side of (\ref{eq: left_right}) has collapsed to a point at
the origin.
If time is switched on, the curve oscillates between these two.
The thin lines shows the same curve, for the same value of $Q'_3$, but
with all intermediate terms zero, i.e. 
$Q_1=\tilde{Q}_1=Q_5=\tilde{Q}_5=0$. These show that a new topological 
charge sector has been found for $\tau_0$.

The next simplest case is $\tau_3$. The conditions 
$Q_1=-\tilde{Q}_5, Q_5= - \tilde{Q}_1$ also mean that the right-hand
side of (\ref{eq: left_right}) is a point. Graphs ($2a$) and ($2b$)
show the two extremal curves. 
For $j=3$ equation (\ref{eq: left_right}) becomes
$$
e^{-x/2}\Big(
1+ \frac 32 e^x\big(2-\cos(\sqrt 3 t)
\big)Q_1^2-Q_3'e^{2x}\Big)\ =\ 0\ ,
$$
and the imaginary part of this is
$$
-\Big(\frac32\Big)^3 p\sin\Big(\frac{41\pi}{42}\Big)e^x\ -\ 
\frac12 e^{2x}\ =\ 0\ .$$
For real $x$, this has no solutions and the curve $\tau_3$ remains in
the lower half plane and does not pass through the origin at any time $t$.
This shows that there is no new sector.

The cases
$\tau_1$ and $\tau_2$ are more delicate. For $\tau_2$, graphs 
($3a$) and ($3b$)
show the ellipse given by the right-hand side of (\ref{eq: left_right})
(in fact this has degenerated to a line), and the two curves given
by the extremal values of $t$. In both cases these curves hit, 
or at least seem to touch, the ellipse. Also note that in ($3b$) the curve
lies in the same position as the standard case 
$Q_1=\tilde{Q}_1=Q_5=\tilde{Q}_5=0$. 
This can also be seen in some of the other graphs.
We shall give two arguments for
demonstrating that this situation is free from zeroes. The
first is slightly more intuitive, but not completely rigorous:
Firstly, we expand $(3a)$, where the curve appears to touch the ellipse,
but we plot the point on the ellipse given by the precise value of $t$,
$\frac{\sqrt{3}}{2}t=\pi$, corresponding  to that curve. 
The expanded graph $(3c)$ shows
that there is no touching, and the point is
below and to the left of the curve, in the way of the motion of the
curve when we switch on $t$. This point is at the tip of the ellipse.
Graph $(3d)$ is a similarly expanded 
graph, at the other extremal value of $t$, that is an expansion
of ($3b$). The point on the ellipse at that value of $t$, 
$\frac{\sqrt{3}}{2}t=\frac{3\pi}{2}$,
is positioned to the left and below the curve, at the origin in fact.
Now the curve oscillates
between these extremal cases, but the point on the ellipse oscillates
at precisely half this frequency. There are two periods in the motion of the
curve during a single period of the point on the ellipse. 
Since this point has moved
half-way down its length during the time the curve
has moved between the extremal curves, and the extremal curve in $(3b)$
intersects the ellipse less than half-way down its length, we can intuitively
infer that the point always stays ahead of the curve throughout the
motion during this
half-period, with the separation along the line increasing.
During the other half-period when the curve moves
back to its starting position, the point on the ellipse moves further
down away from the curve. In the next half-period it moves back
to look like graph $(3d)$, but the point is now travelling up the ellipse,
and the curve is travelling towards its starting position. This is just
the reverse motion of the initial half-period, and is free from a
singularity.  This is not a completely rigorous proof, 
because the motion of the intersection of the curve with the line is not quite
sinusoidal, the tolerances may be such that we still get an intersection
in the first half-period. To disprove this we observe that equation 
(\ref{eq: explicit2}) for $\tau_2$, 
omitting the final term $\omega^{3i}(Q'_3)e^{2x}$,
can be written as (where we must remember that $\omega=e^{2\pi i\over 6}$)
$$1 + i\sqrt{3}\Bigl(Q_1 e^{i{\sqrt{3}\over 2}t} + 
\tilde{Q}_1 e^{-i{\sqrt{3}\over 2}t}\Bigr)e^{\frac{x}2}
- \frac32\Bigl\{Q_1\tilde{Q}_1+\frac12\Bigl(Q_1^2 e^{i\sqrt{3}t} + 
\tilde{Q}_1^2 e^{-i\sqrt{3}t}\Bigr)\Bigr\}e^x, $$
which rather surprisingly is a perfect square,
$$\Bigl(1+i\frac{\sqrt{3}}2(Q_1 e^{i{\sqrt{3}\over 2}t} + 
\tilde{Q}_1 e^{-i{\sqrt{3}\over 2}t})e^{\frac{x}2}\Bigr)^2.$$
Inserting back the final term of (\ref{eq: explicit2}) means that we
must check that the equation
$$1+i\frac{\sqrt{3}}2(Q_1 e^{i{\sqrt{3}\over 2}t} + 
\tilde{Q}_1 e^{-i{\sqrt{3}\over 2}t})e^{\frac{x}2} = 
\pm i\sqrt{Q'_3}e^x,$$
has no solutions, 
or rearranging slightly,
\beq e^{-\frac{x}2}  \mp  i\sqrt{Q'_3}e^{\frac{x}2} = -i\frac{\sqrt{3}}2(Q_1 e^{i{\sqrt{3}\over 2}t} + \tilde{Q}_1 e^{-i{\sqrt{3}\over 2}t})\ .
\label{eq: star}\eeq
Now
$x$ and $t$ have completely decoupled into left and right hand sides of
this equation, and we can therefore plot graphs of the left and right 
hand sides, for both of the possible signs. These are shown in $(4a)$ and
$(4b)$. There are no intersections and therefore $\tau_2$ is free from
zeroes, and the winding number is unchanged.

We treat $\tau_1$ in a similar fashion, graphs ($5a$) and ($5b$)
show the ellipse in comparison to the curve.
It is also possible to show by completing the square, in a similar way
to $\tau_2$,
 that $\tau_1$ is free from zeroes.

A brief calculation will show that
$\tau_4(x,t)=\tau_2(x,t+2\pi/\sqrt 3)$ and 
$\tau_5(x,t)=\tau_1(x,t+2\pi/\sqrt 3)$. Using the results for $\tau_1$ and
$\tau_2$, it is clear that  $\tau_4$ and $\tau_5$
are never zero for any value of $x$ and $t$, and that 
the winding numbers are
unchanged.
\psbild{ht}{(1a)}{9cm}{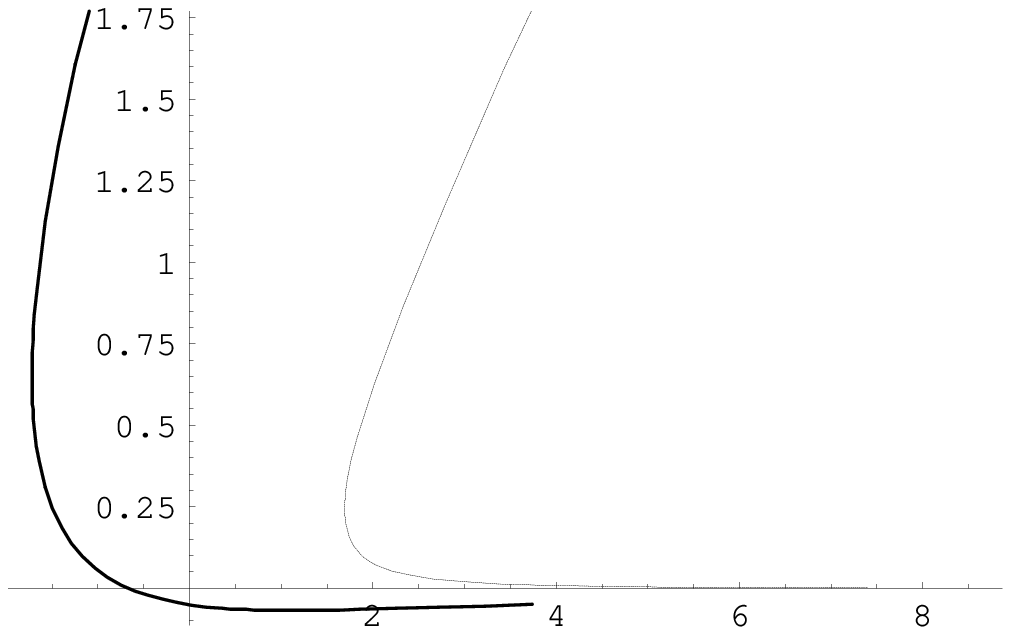}{Eqn. (\ref{eq: left_right}) for $\tau_0$ 
at extremal value of $t$, 
thin curve is case $Q_1=\tilde{Q}_1=Q_5=\tilde{Q}_5=0$}
\psbild{ht}{(1b)}{9cm}{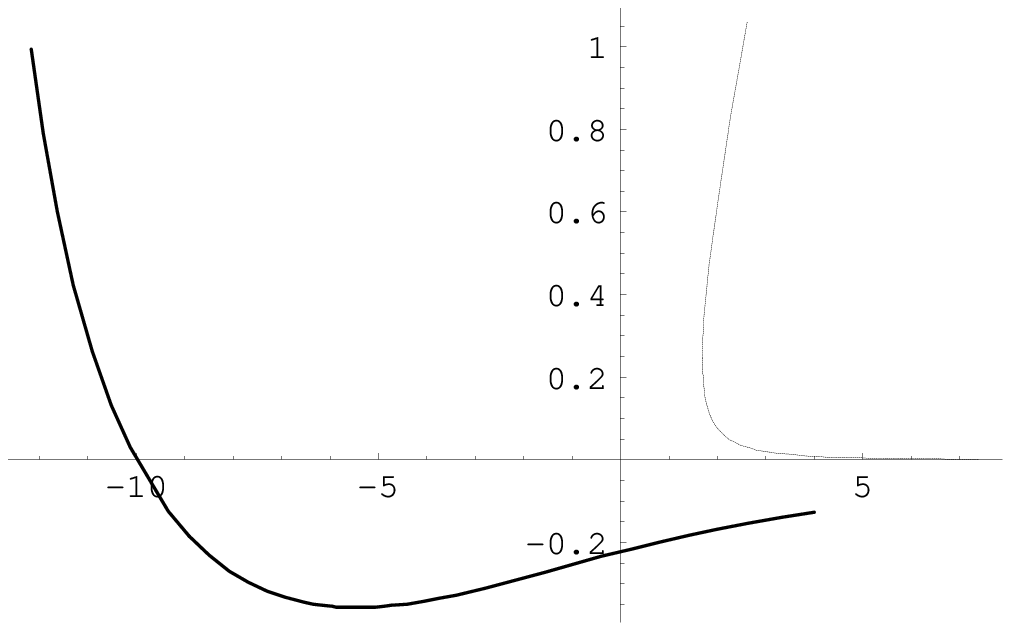}{Eqn. (\ref{eq: left_right}) for $\tau_0$ 
at other extremal value of $t$}
\eject
\psbild{ht}{(2a)}{9cm}{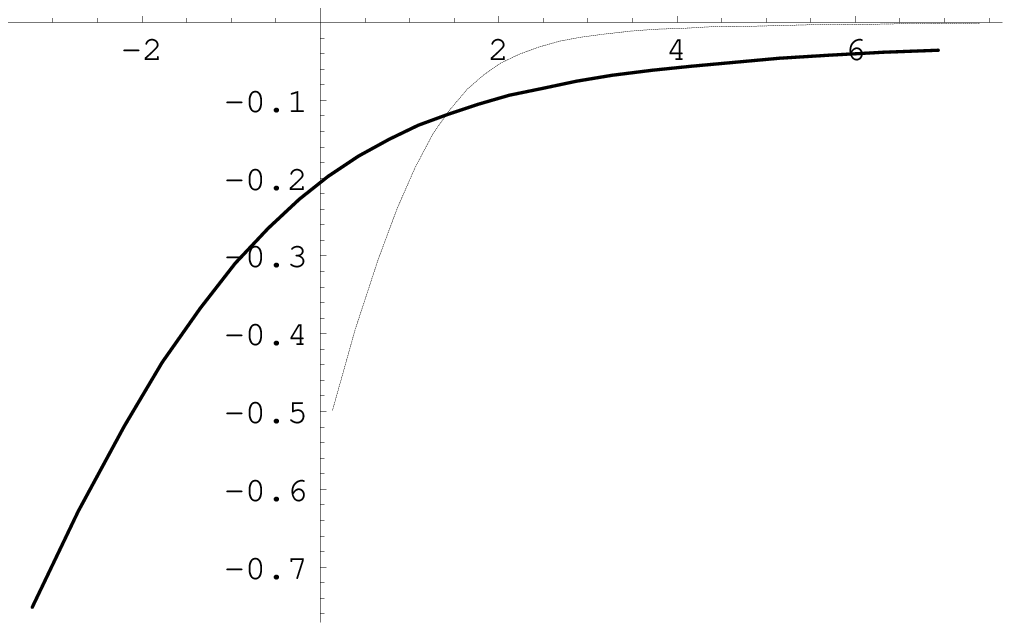}{Eqn. (\ref{eq: left_right}) for $\tau_3$ 
at extremal value of $t$}
\psbild{ht}{(2b)}{9cm}{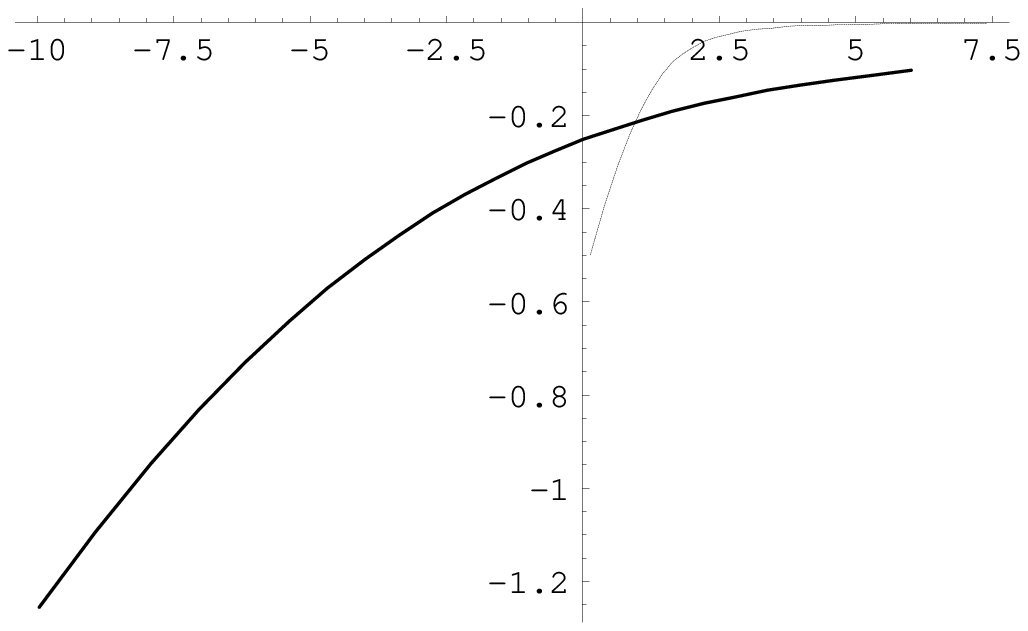}{Eqn. (\ref{eq: left_right}) for $\tau_3$ 
at other extremal value of $t$}
\eject
\psbild{ht}{(3a)}{9cm}{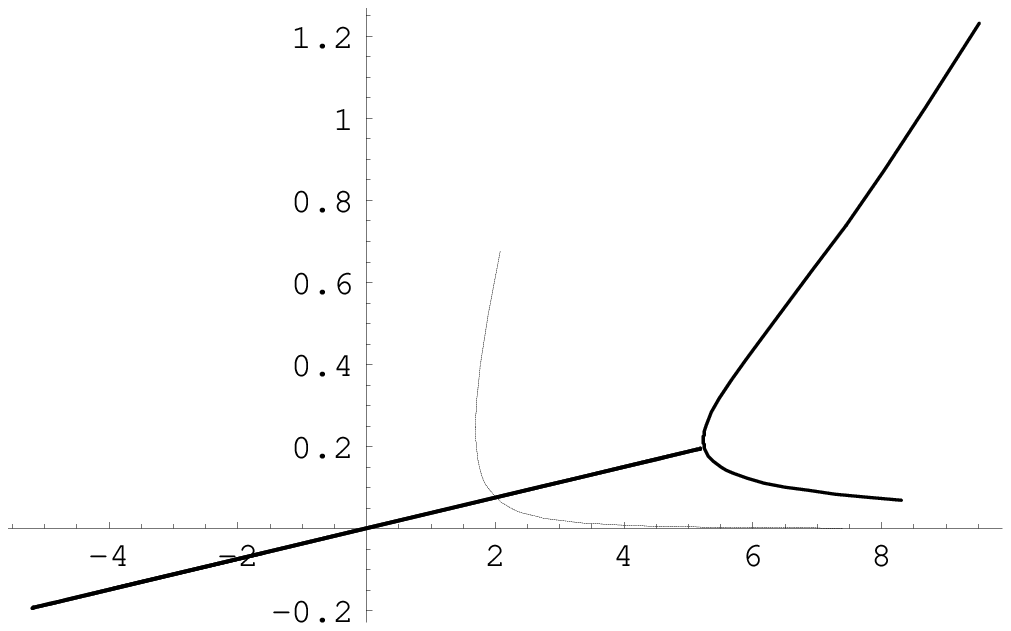}{Eqn. (\ref{eq: left_right}) for $\tau_2$ 
at extremal value of $t$}
\psbild{ht}{(3b)}{9cm}{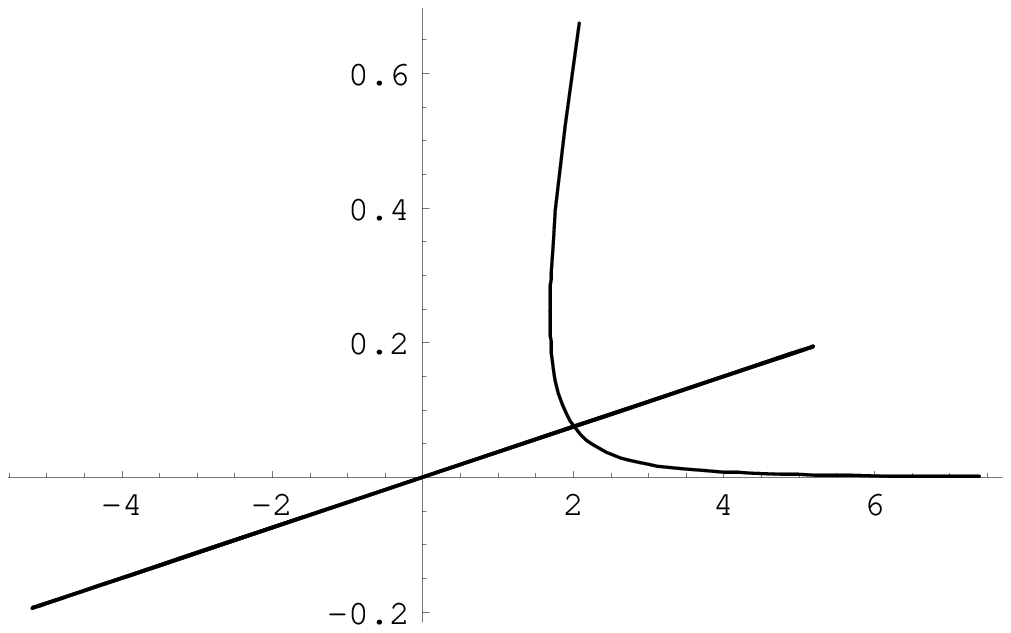}{Eqn. (\ref{eq: left_right}) for $\tau_2$ 
at other extremal value of $t$}
\eject
\psbild{ht}{(3c)}{9cm}{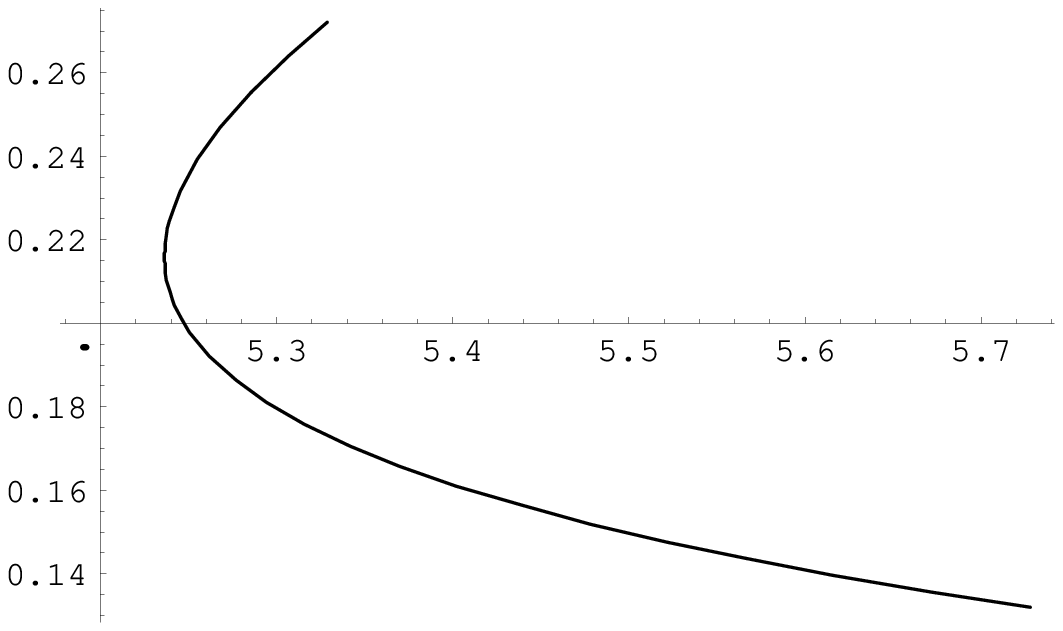}{Enlargement of Fig. ($3a$), point on ellipse
shown for that value of $t$}
\psbild{ht}{(3d)}{9cm}{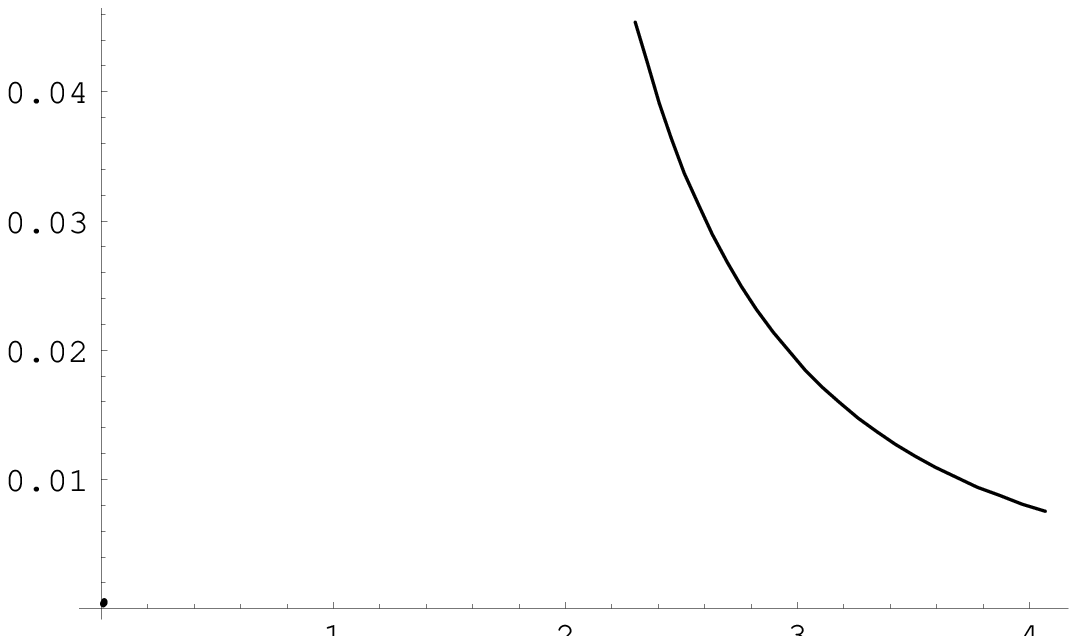}{Enlargement of Fig. ($3b$), point on ellipse
shown for that value of $t$}
\eject
\psbild{ht}{(4a)}{9cm}{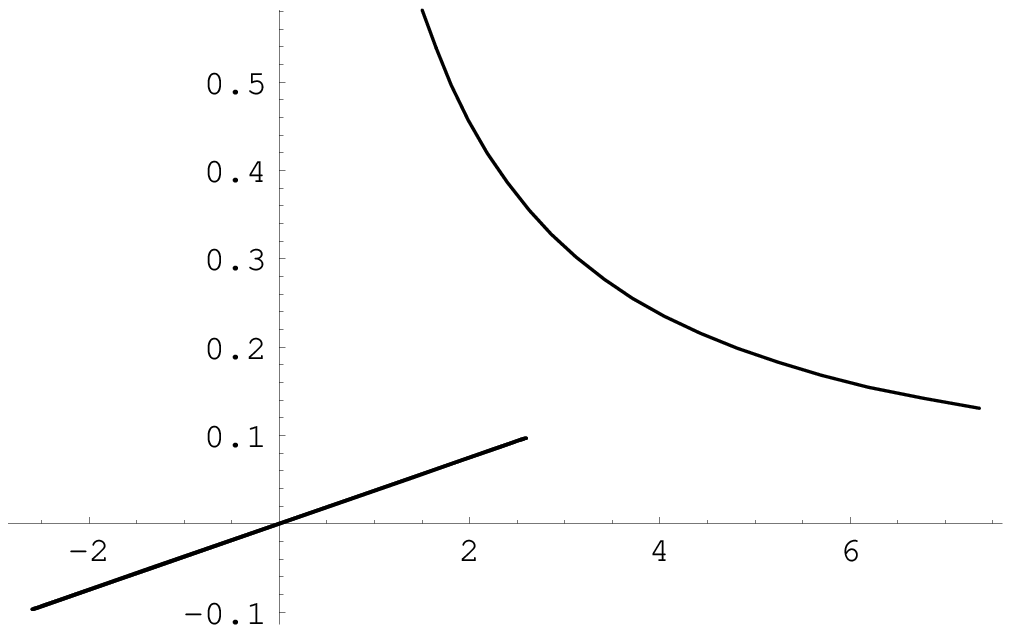}{Eqn. (\ref{eq: star}), plus sign}
\psbild{ht}{(4b)}{9cm}{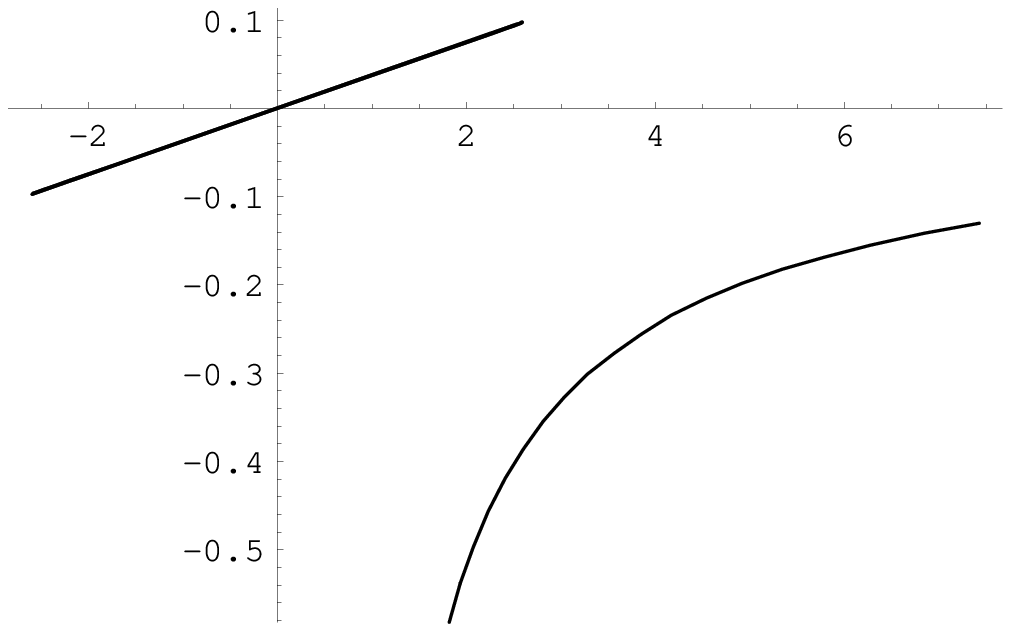}{Eqn. (\ref{eq: star}), minus sign}
\eject
\psbild{ht}{(5a)}{9cm}{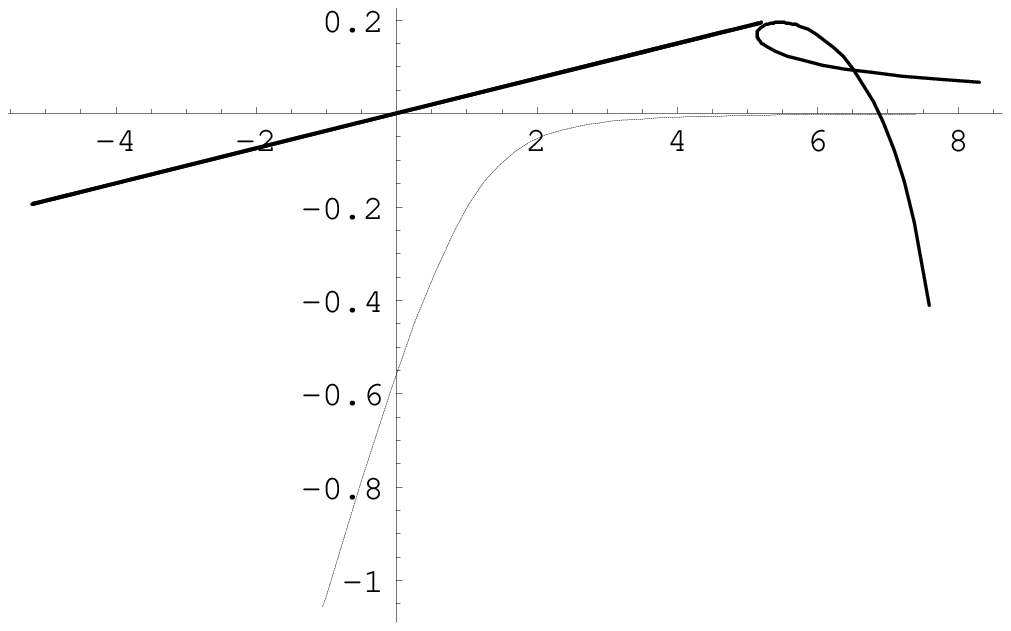}{Eqn. (\ref{eq: left_right}) for $\tau_1$ 
at extremal value of $t$}
\psbild{ht}{(5b)}{9cm}{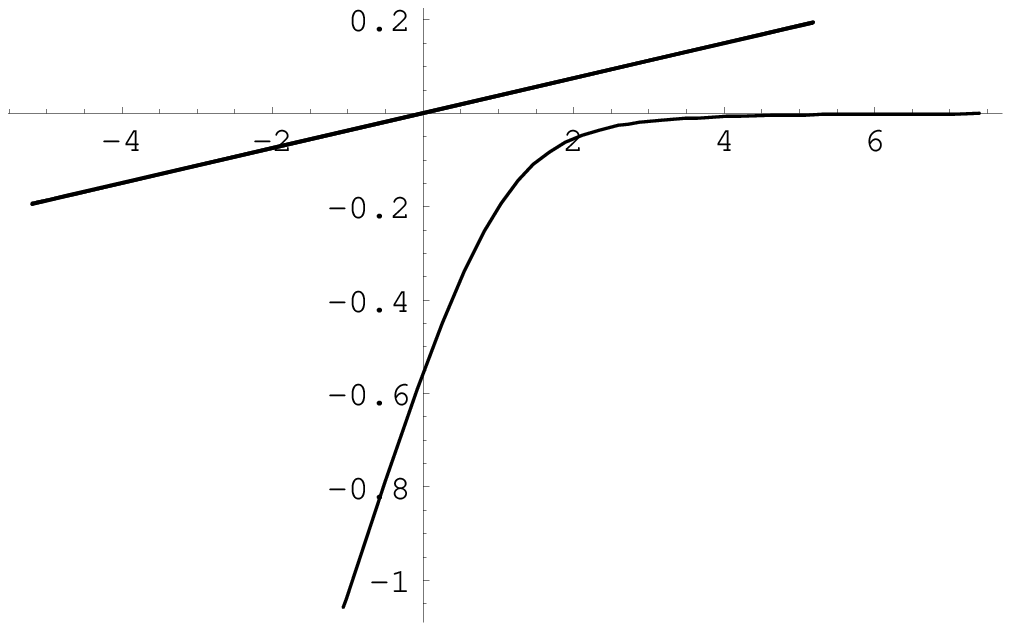}{Eqn. (\ref{eq: left_right}) for $\tau_1$ 
at other extremal value of $t$}
We now discuss the actual values of the topological charges which have been
obtained.
We note that for the given value of $Q'_3$, and 
$Q_1=\tilde{Q}_1=Q_5=\tilde{Q}_5=0$,
we have the topological charge \cite{McGhee}, where
$\alpha_i$ are the simple roots: 
$$\Delta=-\frac12\alpha_1  - \frac12 \alpha_3 
-\frac12\alpha_5.$$  

We have seen how at the special values of $Q$, $\tau_0$ has been tuned into
a new sector, but the sectors of $\tau_j: j=1,\ldots,5$, 
are unchanged in comparison with the thin lines, the case where 
$Q_1=\tilde{Q}_1=Q_5=\tilde{Q}_5=0$.
A change in winding number by $n\in{\Bbb Z}$
 of
$\tau_j$ has the effect of changing the charge by $n\alpha_j$, with
the natural understanding that $\alpha_0=-\sum_{i=1}^5\alpha_i$.

This gives rise to the topological charge:
$$\Delta=\frac12\alpha_1 + \alpha_2 + \frac12 \alpha_3 
+\alpha_4 +\frac12\alpha_5.$$ 

The analysis can be repeated by focussing on the other $\tau_j$, 
$j=1,\ldots,5$, with 
$$Q_1=-\omega^{4j}\tilde{Q}_5, Q_5= - \omega^{-4j}\tilde{Q}_1.$$
If $j$ is even we can pick  $Q_1$  in terms of the old $Q_1$ as 
$\omega^{-j}Q_1$, and   $\tilde{Q}_1$ as $\omega^{-j}\tilde{Q_1}$.
It is easy to see that the only
difference from the previous case is a permutation of the tau functions,
and therefore the other $\tau$'s must be free from zeroes and the 
 winding numbers are unchanged. On the other hand, if $j$ is odd, we must
pick a different $Q_1$ and $\tilde{Q}_1$ unrelated to  before. 
It is possible to do this, and the analysis for showing 
that $\tau_j$'s winding number is
changed, and the others are free from zeroes and their winding numbers
unchanged, is essentially identical to the $j$ even case.

Treating $\tau_1$ gives a new topological charge.
Hence we find the charge
$$\Delta=\frac12\alpha_1 -  \frac12 \alpha_3 
-\frac12\alpha_5.$$

Treating $\tau_2$ gives 
$$\Delta=-\frac12\alpha_1 - \alpha_2 - \frac12 \alpha_3 
- \frac12\alpha_5.$$  

Treating $\tau_3$ gives:
$$\Delta=-\frac12\alpha_1 + \frac12 \alpha_3 
-\frac12\alpha_5.$$  

Treating $\tau_4$ gives:
$$\Delta=-\frac12\alpha_1 - \frac12 \alpha_3 
- \alpha_4 - \frac12\alpha_5.$$  

Treating $\tau_5$ gives:
$$\Delta=-\frac12\alpha_1 - \frac12 \alpha_3 
+\frac12\alpha_5.$$  

This gives $6$ new charges which are weights of the fundamental representation
$V_3$. We can get $6$ more by negating all these. These arise because we
can pick the topological charge of the standard solution, when 
$Q_1=\tilde{Q}_1=Q_5=\tilde{Q}_5=0$, to be the other of the two
possibilities, when $Q'_3$ is moved through the real axis. It is then
possible to find suitable 
$Q_1,\tilde{Q}_1,Q_5,\tilde{Q}_5$, so that this statement is true.

Therefore there are now $12+2=14$ topological
charges of the fundamental representation $V_3$ filled by classical
soliton solutions. These do not include the highest weight of $V_3$.
As $V_3$ is $20$ dimensional,
 there are $6$ as yet unfilled ones. It is curious that $6$
is precisely the number of $\tau$ functions in this model, which suggests 
that the charges can be found by similar methods, since the tuning method
described above shows that new topological charges can be found in sets
of $6$. 

We also remark that the first attempt at analysing the solution
(\ref{eq: explicit2}) was to consider $Q_5=\tilde{Q}_5=0$, where the
time dependence in the intermediate term in curly brackets of 
(\ref{eq: left_right}) does not appear. In order to tune the curve for 
$\tau_0$, say, past
the ellipse given by the right-hand side, it is necessary to   
choose quite large values for the parameters $Q_1$, $\tilde{Q}_1$ 
(in comparison with the values actually chosen), and
to choose the phase of $Q_1\tilde{Q}_1$ within a narrow range of acceptable
values. These choices mean that at least one of the other tau functions
obviously hits zero, with no possibility of fine adjustment of the
$Q$'s. Therefore the case considered here is indeed the simplest.
 
We conclude by observing that the time delay of this new 
soliton when it interacts with a standard soliton of a different, or
the same, species, given by (\ref{eq: standard_one}), is the
same as that of standard solitons interacting with each other.
This is because the $W$ with the highest exponential behaviour in $x$,
for each of the two solitons, determines the time delay, see \cite{FJKO}.
The $X$ factor which arises when these two $W$'s combine in the 
multi-soliton solution (see the rules at the beginning of section 5) is
essentially the exponential of the time delay.  This $X$ factor is the
same regardless of whether there are intermediate terms in the tau
functions: these are neglegible asymptotically, and do not
contribute when the time delay is calculated. 
This argument is equally valid for the new solitons 
interacting with each other. Therefore the new solitons are exactly what
are required to fill weights in the fundamental representations.

\noindent {\bf Acknowledgements}

P.R.J. is very happy to thank V. Pasquier for discussions during
the preliminary stages of this work. The matrix inversion,
equation (\ref{eq: inversion}),
was calculated using REDUCE. 
P.R.J. also acknowledges financial support from the European Union 
TMR network programme, contract ERBFMRXCT960012, held at CEA-Saclay and
the Albert Einstein Institut, Potsdam.

\end{document}